\documentclass[a4paper, 11pt]{article}    % Specifies the document style.
\usepackage{latexsym}
\usepackage{amssymb}
\usepackage{amsmath}
\usepackage{amsfonts}
\usepackage{bbm}
\usepackage{graphicx}
\usepackage{float}
\usepackage[english]{babel}
\usepackage{multirow}
\usepackage{float}
\usepackage{soul,color}
\restylefloat{table}
\usepackage{caption}
\usepackage{placeins}
\usepackage{hyperref}
\usepackage{wrapfig}
\usepackage{cleveref}
\usepackage{enumitem}
\usepackage[toc,page]{appendix}
\usepackage[normalem]{ulem}
\useunder{\uline}{\ul}{}
\usepackage{ltxtable}
\usepackage{subfig}

\newcolumntype{L}[1]{>{\raggedright\let\newline\\\arraybackslash\hspace{0pt}}m{#1}}
\newcolumntype{C}[1]{>{\centering\let\newline\\\arraybackslash\hspace{0pt}}m{#1}}

\newcommand{\quota}[1]{``#1''}
\makeatletter
\def\@maketitle{
\begin{center}
{\Huge \bfseries \sffamily \@title }\\[4ex]
{\Large  \@author}\\[2ex]
\includegraphics[width = 30mm]{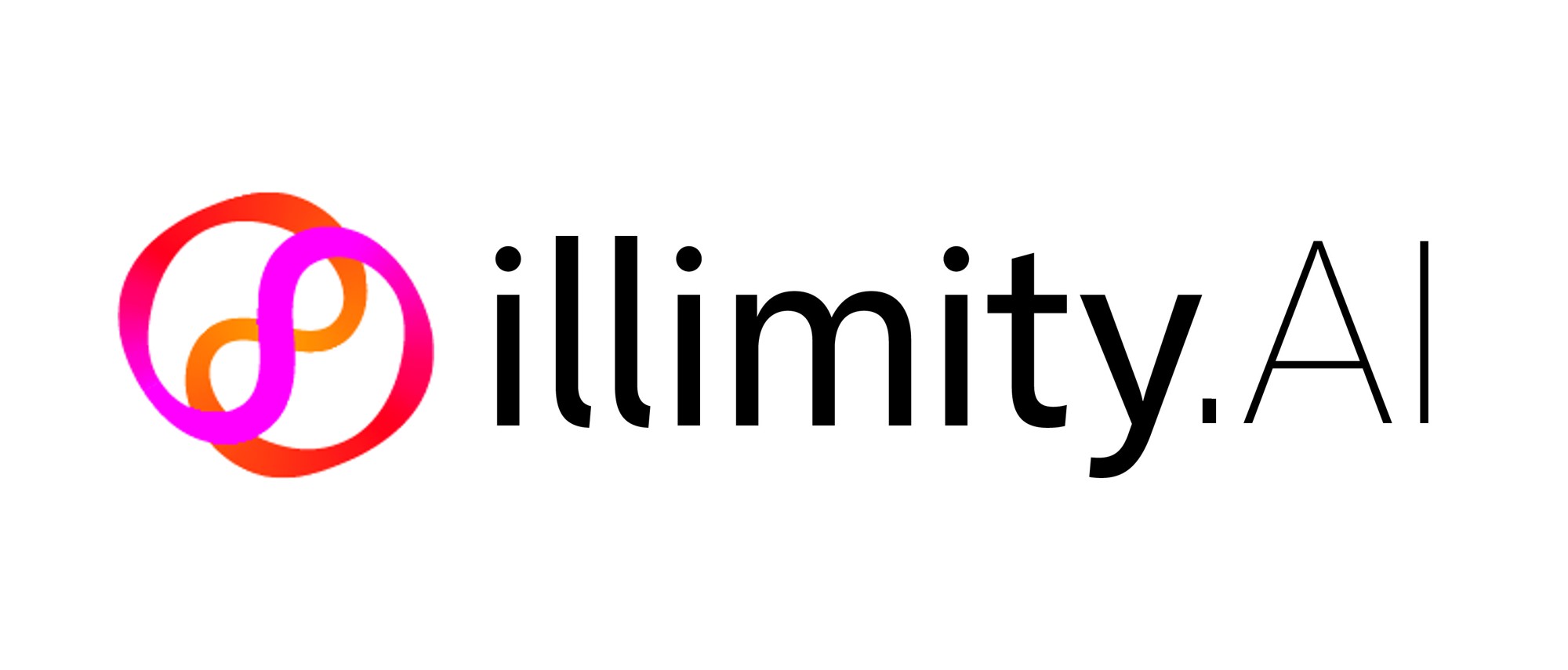}\\[4ex]
\@date\\[4ex]
\end{center}}
\makeatother
\title{ \bf Machine Learning approach for Credit Scoring}  % Declares the document's title.
\author{A. R. Provenzano\footnote{corresponding author: angela.provenzano@illimity.com}\,\,, D. Trifir\`o, A. Datteo, L. Giada, N. Jean, A. Riciputi, G. Le Pera, M. Spadaccino, L. Massaron and C. Nordio}
\begin{document}           % End of preamble and beginning of text.
\maketitle                 % Produces the title.

\begin{center}
\tt \Large Working paper\footnote{This paper reflects the authors' opinions and not necessarily those of their employers.}
\end{center}

\vspace{10mm}

\begin{abstract}
\it \noindent
In this work we build a stack of machine learning models aimed at composing a state-of-the-art credit rating and default prediction system, obtaining excellent out-of-sample performances. Our approach is an excursion through the most recent ML / AI concepts, starting from natural language processes (NLP) applied to economic sectors' (textual) descriptions using embedding and autoencoders (AE), going through the classification of defaultable firms on the base of a wide range of economic features using gradient boosting machines (GBM) and calibrating their probabilities paying due attention to the treatment of unbalanced samples. Finally we assign credit ratings through genetic algorithms (differential evolution, DE). Model interpretability is achieved by implementing recent techniques such as SHAP and LIME, which explain predictions locally in features' space.
\end{abstract}

\bigskip
{ \bf JEL} Classification codes: C45, C55, G24, G32, G33
{ \bf AMS} Classification codes: 62M45, 68T01, 68T50, 91G40
\bigskip

{\bf Keywords:} Artificial Intelligence, Machine Learning, Explainable AI, Autoencoders, Embedding, LightGBM, Differential Evolution, SHAP, LIME, Credit Risk, Rating Model, Default, Probability of Default, Classification

%%%%%%%%%%%%%%%%%%%%%%%%%%%%%%%%%%%%%%%%%%%%%%%%%%%%%%%%%%%%%%%%%%%%%%%%%%%%%%%%%%%%%%%%%

\section*{Introduction}
\label{section:intro}

In the aftermath of the economic crisis, the probability of default (PD) has become a topical theme in the field of financial research. Indeed, given its usage in the risk management, in the valuation of the credit derivatives, in the estimation of the creditworthiness of a borrower and in the calculation of economic or regulatory capital for banking institution (under Basel II), incorrect PD prediction can lead to false valuation of risk, unreasonable rating and incorrect pricing of financial instruments. In the last decades, a growing number of approaches has been developed to model the credit quality of a company, by exploring statistical techniques. Several works have employed probit models\cite{mizen2012forecasting} or linear and logistic regression to estimate company ratings using the main financial indicators as model input. However, these models suffer from their clear inability to capture non-linear dynamics, which are prevalent in financial ratio data\cite{gurny2013comparison}. New statistical techniques, especially from the field of machine learning, have gained a worldwide reputation thanks to their ability to efficiently capture information from big dataset by recognizing non-linear patterns and temporal dependencies among data. Zhao et al. (2015) \cite{zhao2015investigation} employed feed forward neural networks in credit corporate rating determination. Petropopulos et al.\cite{petropoulos2019robust} explore two state of the art techniques namely Extreme Gradient Boosting (XGBoost) and deep learning neural networks in order to estimate loan PD and calibrate an internal rating system, useful both for internal usage and regulatory scope. Addo et al. (2018)\cite{addo2018credit} built binary classifiers based on machine and deep learning models on real data to predict loan probability of default. They observed that the tree-based models are more stable than ones based on multilayer artificial neural networks.

Starting from these studies, we propose a sophisticated framework of machine learning models which, on the basis of company annual (end-of-year) financial statements coupled with relevant macroeconomic indicators, attempts to classify the status of a company (performing - \quota{in-bonis} - or defaulted) and to build a robust rating system in which each rating class will be matched to an internally calibrated default probability. In this regard, here the target variable is different from a previous work by some of the authors \cite{provenzano2019artificial}, where the goal was to predict the credit rating that Moody's would assign, according to an approach commonly called \quota{shadow rating}. The novelty of our approach lies in the combination of data preprocessing algorithms, responsible for feature engineering and feature selection, and a core model architecture made of a concatenation of a Boosted Tree default classifier, a probability calibrator and a rating attribution system based on genetic algorithm. Great attention is then given to model interpretability, as we propose two intuitive approaches to interpret the model output by exploring the property of local explainability. In details, the article is composed of the following sections: \Cref{section:data} is devoted to describe the input dataset and the preprocessing phase; \Cref{section:model} in which the core model architecture is explained; \Cref{section:results} which collects results from the core model structure (i.e. default classifier, PD calibrator and rating clustering); finally \Cref{section:explain} is left to model explainability.

%%%%%%%%%%%%%%%%%%%%%%%%%%%%%%%%%%%%%%%%%%%%%%%%%%%%%%%%%%%%%%%%%%%%%%%%%%%%%%%%%%%%%%%%%

\section{Dataset description}
\label{section:data}

\begin{figure}[h!]
     \centering
     %\vspace{10pt}
	\includegraphics[scale=0.6]{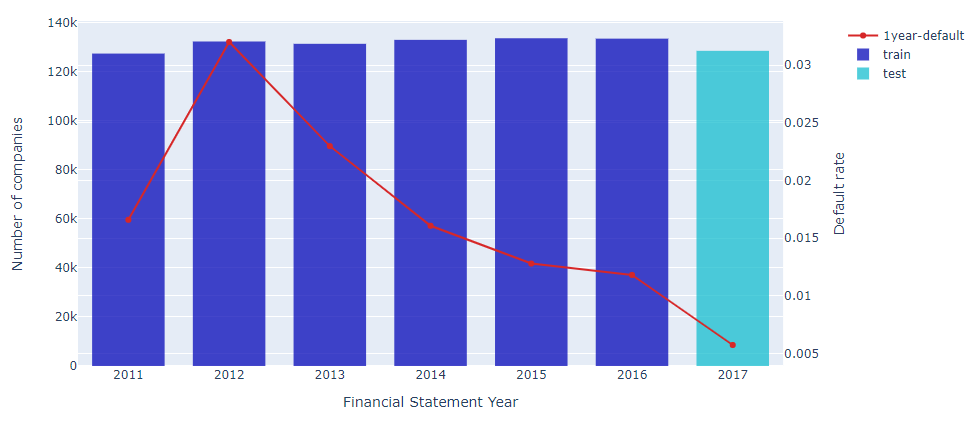}
     \caption{Number of balance-sheets per financial statement year and the corresponding $1$-year default rate. The default rate shows an increasing trend in the $2011-2012$ period, and then decreases till $2017$. }
     \label{fig:histo}
\end{figure}

Data used for model training have been collected from the Credit Research Database (CRD) provided by Moody's, and consist of $919,636$ annual (end of year) financial statements of $157,986$ Italian companies belonging to different sectors (e.g. automotive, construction, consumer goods durable and not-durable, energy, high-teach industries, media, etc. to the exclusion of FIRE sector, i.e. finance, insurance, and real estate sectors). The dependent variable in our dataset, i.e. the \emph{target} of the proposed default prediction model, is a binary indicator with the value of $1$ flagging a default event (i.e. a bankruptcy occurrence over a one-year horizon), $0$ otherwise. \\In accordance to the above-defined target variable, input variables of our model have been selected to be consistent with factors that can affect the companies capacity to service external debt (a full explanation of the input model's features is reported in Appendix \autoref{appendix}). In particular they consist of balance-sheet indexes and ratios, and \emph{Key Performance Indicators} (KPI) calculated from CRDs financial reports\cite{CRD}. The latter include indicators for efficiency (i.e. measures of operating performance), liquidity (i.e. ratios used to determine how quickly a company can turn its assets into cash if it is experiencing financial distress or impending bankruptcy), solvency (i.e. ratios that depict how much a company relies upon its debt to fund operations) and profitability (i.e. measures that demonstrate how profitable a company is).
Since business cycles can have great impact on a firm profitability and influence its risk profile, we joined original information with more general macro variables ($2$ years lagged historical data) addressing the surrounding climate in which companies operate. Among the wide range of macroeconomic indicators provided by Oxford Economics \cite{oxford}, a subset of the most influential ones has been selected as explanatory variables\footnote{The list of selected indicators are reported in \Cref{appendix: macro}}. Some of them are country-specific, others are common to the whole Eurozone\footnote{Regional aggregate \emph{Eurozone} includes the following countries: Austria, Belgium, Cyprus, Estonia, Finland, France, Germany, Greece, Ireland, Italy, Latvia, Lithuania, Luxembourg, Malta, Netherlands, Portugal, Slovenia, Slovakia and  Spain}. The combined dataset of balance-sheet indexes, financial ratios and macro variables along with data transformations and feature selection (better described hereafter in \Cref{subsection:features}), led to a set of 179 features and covers the period $2011-2017$. \\As fully described hereafter, the obtained dataset was split into three parts: an out-of-time dataset which includes the data referred to year $2017$ (marked in light-blue in \autoref{fig:histo}), used to test model performance; a stratified pair of train/test dataset (constituting each the $80\%$ and the $20\%$ of the input dataset) which covers the period $2011-2016$, employed for model development and calibration.

%\vspace{10pt}
\FloatBarrier

%and in order to clean up data from possibly erroneous observations and create a robust and reliable dataset, we perform anomaly detection\footnote{Anomaly detection refers to identification of items or events that do not conform to an expected pattern or to other items in a dataset.} on financial statement years via Isolation Forest Algorithm provided by sklean. In particular, Isolation Forest (IForest) is an ensemble regressor which "\emph{isolates}" outliers by randomly selecting a feature and then randomly selecting a split value between the maximum and minimum values of the selected feature \cite{liu2008isolation}. In details, since recursive partitioning can be represented by a tree structure, the number of splitting required to isolate a sample is equivalent to the path length from the root node to the terminating node. Random partitioning produces noticeably shorter paths for anomalies. Hence, when a forest of random trees collectively produce shorter path lengths for particular samples, they are highly likely to be anomalies. As a result of anomaly detection, observations coming from $2008$ and $2018$ have been recognized as anomalous, since the anomaly ratio is near to $1$, as expected for outliers.

\subsection{Feature engineering and feature selection}
\label{subsection:features}

A preliminary step for building a machine learning model consists in generating a set of features suitable for model training. This task involves data manipulation processes like transformation of categorical features, missing values treatment, infinite values handling, outliers detection, data leakage avoidance. In particular, categorical, non-ordinal variables are one of the main issue that must be tackled in order to feed any machine learning model\cite{zheng2018feature}.\\Different \emph{encoding} technique can be used to make categorical data legible for a machine learning algorithm. The most common way to deal with categories is to simply map each category with a number. By applying such transformation, known as \emph{Label Encoding}, a model would treat categories as ordered integers, which would imply non-existent ordinal relationships between data, that could be misleading for model training. Another simple way to handle categorical data is the \emph{One-Hot Encoding} technique, which consists in transforming each categorical feature into fixed-size sparse vector of all zeros but $1$ in the cell used uniquely to identify specific realization of that variable. The main drawback of this technique relies on the fact that categories with an high number of possible realizations would generate large dimension datasets, which makes it a memory-inefficient encoder. Moreover this sparse representation does not preserve similarity between feature values. An alternative approach to overcome these issues is represented by \emph{Categorical Embedding}, which consists in mapping via Deep Neural Network (DNN) each possible discrete value of a given categorical variable in a low-dimensional, learned, continuous vector representation. This method allows to place each categorical feature in a Euclidean space, keeping coherent relationship with other realization of the same variable. The extension of categorical embedding approach to words and document representation is known as \emph{Word Embedding}\cite{guo2016entity}. In particular, \emph{Sentence Embedding} is an application of word embedding aiming at representing a full sentence into a vector space. In this study, we applied sentence embedding to represent industry sectors descriptions associated to each "NACE code"\footnote{The "\emph{Statistical Classification of Economic Activities in the European Community}", commonly referred to as NACE, is the industry standard classification system used in the European Union}. In order to guarantee the "\emph{semantics}" of the original data and work in a low dimensional space, we propose a framework of embedding with autoencoder regularization, in which the original data are embedded into low dimension vectors. The obtained embeddings maintain local similarity and can be easily reverted to their original forms. The encoding of the NACE is a novel way to overcome the NACE$1$-NACE$2$ mapping conundrum. In our dataset both NACE are used and as already stated in many papers the two encoding systems are not fully compatible\cite{perani2015matching}. Moreover the NACE encoding allows for proper industry segment description of multi sector firms that cannot be easily described by a single NACE code, further extending the predictive power of the economic sector category. \\A different encoding method is the \emph{Target encoding}, in which categorical features are replaced with the mean target value for samples having that category. This allows to encode an arbitrary number of features without increasing data dimensionality. However, as a drawback, a naive application of this type of encoding can allow data leakage, leading to model overfitting and poor predictive performance. A target encoding algorithm developed for preventing data leakage is know as \emph{James-Stein estimator}, and is the one used in our model. In more details, it transforms each categorical feature with a weighted average of the mean target value for the observed feature value and the mean target value computed regardless of the feature realization.

As described above, some feature transformations could result in a general increase of input data dimensionality, which makes it urgent to implement a robust and independent feature selection framework. In fact, training a machine learning model to a huge number of independent variables is doomed to suffer from the so-called \emph{curse of dimensionality}\cite{bellman1957dynamic}, i.e. the problem of exponential increase in volume associated with adding extra dimensions to a vector space. We employed a voting ensemble of models to independently assign importance to the available features and efficiently select those features which will contribute most to model prediction.

Hereafter in this section we will look into the implementation of satellite models aiming at: performing sentence embedding of the industry sectors description; reducing embeddings dimensionality via stacked autoencoder; selecting relevant features via voting approach.

\paragraph{Sentence Embedding of sector descriptions}

A common practice in Natural Language Processing (NLP) is the use of pre-trained embeddings to represent words or sentence in a document. Following this common practice, we use the pre-trained models built in SpaCy NLP library for embedding the sequence of NACE sector textual description. In particular, we performed sentence embedding i.e. we transform each description into a $300$-dimensional real-value vector. Each sentence embedding is automatically constructed by SpaCy averaging the $300$-dimensional real-value pre-trained vectors which map each word in that sentence. \\Here's a glimpse at how spaCy processes textual data. It first segments text into words, punctuations, symbols and others by applying specific rules to each language (i.e. it tokenizes the text). Then it performs Part-of-speech (POS) tagging to understand the grammatical properties of each word by means of built in statistical model. A model consists of binary data trained on a dataset large enough to allow the system to make predictions that generalize across the language. A key assumption to the word embedding approach is the idea of using for each word a dense distributed representation learned based on the usage of words \cite{bengio2003neural}. This allows words that are used in similar ways to have similar representations, naturally capturing their meaning\cite{goldberg2017neural}. Given the high importance industry sector has in financial literature as default prediction driver, embedding NACE industry descriptions improves the overall model performance in application by helping the model to generalize better and to smoothly handle unseen elements.

\paragraph{Dimensionality reduction via stacked autoencoder} %DA MODIFICARE

The aforementioned world embedding models are powerful way to represent categorical variables which preserve relationship between data, but at the cost of increase in dimensionality. In order to reduce the number of dimensions of the output embeddings from $300$ to $5$, a \emph{stacked autoencoder} (SAE) of $6$ layers\footnote{a $3$-layer encoder and $3$-layer decoder} has been developed via tensorflow\cite{geron2017hands}. \\In details, \emph{autoencoders} (AE) are a family of neural networks in which input and output coincide. They work by compressing the input into a latent-space representation and then reconstruct the output by means of this representation. They consist of two principal component: the \emph{encoder} which takes the input and compresses it into a representation with less dimensions, and the \emph{decoder} which tries to reconstruct the input. Among AEs, stacked autoencoders are deep neural networks in which the output of each hidden layer is connected to the input of the successive hidden layer. All hidden layers are trained by an unsupervised algorithm and then fine-tuned by a supervised method aimed at minimize the cost function. Since they can learn even non-linear transformations, unlike PCA, by using a non-linear activation function and a multiple layer structure, autoencoders are efficient tools for dimensionality reduction. Moreover, in our application, SAE exhibited a low reconstruction loss\footnote{The \emph{reconstruction loss} is the loss function (usually either the mean-squared error or cross-entropy between the reconstructed output and the input) which penalizes the network for creating outputs different from the original input} (around $6\%$ of MSE), contrary to a low fraction of variance explained by the PCA.

\paragraph{Voting approach for feature selection}

Feature selection is a key component when building machine learning models. We can either demand this task to the main model or use a set of lighter models in a preparatory task so that the required effort for further feature selections will be reduced when training the main model. This is particularly useful for multi parameters models like Light-GBM where the training phase involves also the calibration of a set of hyperparamters usually spanning very wide ranges. Neglecting the expert based component, algorithmic feature selection methods are usually divided into three classes: \emph{filter} methods, \emph{wrapper} methods and \emph{embedded} methods.
Filter-based methods apply a statistical measure to assign a score to each feature; variables of the starting dataset are then ranked according to their scores and either selected to be kept or removed. Wrapper-based methods consider the selection of a set of features as a "search problem", where different combinations are prepared, evaluated and compared to other combinations. In details, a predictive model is used to evaluate a combination of features and assign a score based on model accuracy. Embedded-based methods learn which features best contribute to the accuracy of the model while the model is being created. \\We combined a set of $6$ different models for feature selection, stacking each algorithm into a hard voting framework where features which receive the highest number of votes among all the models have been selected. In particular, after having transformed categorical features via target encoding (by means of James-Stein encoder), each feature in the dataset has been ranked on the basis of the following models:

\begin{itemize}
\item \textbf{Pearson criterion}. It is a \emph{filter-based} method which consists in checking the absolute value of the Pearson correlation between the target and features in the input dataset and keeping the top $n$ features based on this score.

\item \textbf{Chi-squared criterion}. It is another \emph{filter-based} method in which we calculate the chi-squared metric between each feature and the target and select the desired number of features which exhibit the best chi-squared scores. The underlying intuition is that if a feature is independent to the target it is uninformative for classifying information.

\item \textbf{Recursive Feature Elimination} (\textbf{RFE}). This is a \emph{wrapper-based} method whose goal is to select features by recursively considering smaller and smaller sets of features. First, the estimator is trained on the initial set of features and the importance of each feature is computed. In our specific case the estimator used is a Logistic Regression. Then, the least important features are pruned from current set of features. That procedure is recursively repeated on the pruned set until the desired number of features to select is eventually reached.

\item \textbf{Random Forest Classifier} (\textbf{RF}\footnote{RF is an \emph{ensemble} of Decision Trees generally trained via \emph{bagging method}: this approach consists in using the same training algorithm for every predictor, but training them on different random subsets of the Train set. Once all predictors are trained, the ensemble can make a prediction for a new instance by simply aggregating the predictions of all predictors}). This is a \emph{wrapper-based} method that uses a built-in algorithm for feature selection. In particular, variables are selected accordingly to feature importance, obtained by averaging of all decision tree feature importance.

\item \textbf{Logistic Lasso Regression} It is an \emph{embedded-based} method which uses the built-in feature selection algorithm embedded into the Logistic Regression with L1 regularization model.

\item \textbf{Light-GBM}\cite{LGBM} (\textbf{LGBM}\footnote{LGBM is a fast, high-performance gradient boosting framework based on decision tree algorithm}). It is a \emph{wrapper-based} method analogous to the above-mentioned RF classifier.
\end{itemize}

\section{Model architecture}
\label{section:model}

Moving beyond the satellite models described in \Cref{subsection:features}, and used in the preprocessor phase, in this section we will present the core model architecture. It consists of a concatenation of three machine learning models aiming at building a robust and reliable framework whose purpose is not only to classify the company status (as in bonis or defaulted), but also to construct an internally calibrated rating system in which each rating class will correspond to a self-consistent default probability. \\In details, we developed a Boosted Tree algorithm in order to classify the status of a company over a one-year horizon. This is a classical binary classification problem whose target variable is 1 in case of default event or 0 otherwise.
Hyper-parameters tuning has been performed via an extension of cross validation for time-series that will be further described hereafter. \\The purpose of the second model is to fit the output score of the binary classificator to the actual default rate. We built a calibrator which consists of a Logistic Regression trained on the leaf assignments of the Boosted Tree. \\Finally, we calibrated our own rating system by splitting the refitted default probability into $9$ clusters via genetic algorithm.

\subsection{Light-GBM for default classification}

In order to leverage the availability of a large scale dataset, enriched with a high number of features, we developed a robust machine learning approach based on \emph{Gradient Boosting} decision trees known as \emph{Light-GBM}.\\Gradient Boosting trees model\cite{friedman2002stochastic} is one method of combining a group of  "weak learners" (specifically decision trees) in order to form a "strong predictor" model, by reducing both variance and bias. Differently from other tree methods like Random Forest, Boosted tress work by sequentially adding predictors to an ensemble, each one corrects its predecessor by trying to fit the new predictor to the residuals of the previous one. These residuals are the gradient of the loss functional being minimized, with respect to the model values at each training data point evaluated at the current step. Specifically, at each iteration a sub-sample of the training data is drawn at random (without replacement) from the full training data-set. This randomly selected sub-sample is then used in place of the full sample to fit the "weak learner" and compute the model update for the current iteration. \\In particular \emph{Light GBM} (LGBM) is a fast, high-performance gradient boosting framework based on decision tree algorithm, which has proved to be highly effective in classification and regression models when applied on tabular, structured data, such as the ones we are dealing with. The model hyper-parameters have been tuned via \emph{out-of-time cross-validation} procedure based on a custom extension of the \emph{F$_{\beta}$-measure} where the balance of specificity\footnote{The specificity is defined as the number of true negatives over the number of true negatives plus the number of false negatives} (also called "the true negative rate") and recall\footnote{The recall is defined as the number of true positives over the number of true positives plus the number of false negatives} (also called "the true positive rate") in the calculation of the harmonic mean is controlled by a coefficient called $\beta$ as follows:

\begin{equation}
F_{\beta} = (1+\beta^2)\frac{\text{specificity} \cdot \text{recall}}{\beta^2 \cdot \text{specificity} + \text{recall}}
\end{equation}

In this procedure, each test sets consist of a single year of future observation, while the corresponding training sets are made up of the observations that occurred prior to the observation that forms the test set. In this way, the model is optimized in predicting what will happen in the future using only information available up to the the present day. The objective function used for the classification problem was the \emph{log-loss}, which measures the distance between each predicted probability and the actual class output value by means of a logarithmic penalty. Due to the high unbalance between $0$ and $1$ target flags, we have used the modified unbalanced log-loss, by setting the \emph{scale$\_$pos$\_$weight} parameter of the Light-GBM equal to the ratio between the number of $0$s and the number of $1$s. Other objective functions we have used like Focal Loss\cite{lin2017focal} and custom weighted log losses have not given any specific advantage compared to the unbalanced log loss.

\subsection{Probability calibration for tree-base model}

A natural extension of the issue of classification of corporate default forecast consists in predicting the probability of default. Complex non-linear machine learning algorithms can provide poor estimates of the class probabilities, especially in case the target variable is highly unbalanced, so that the distribution and behaviour of the probabilities may not be reflective of the true underlying probability of the sample. The unbalanced log loss objective chosen for the classification task creates a custom metric in the default probability space that is reflected in distorted class probabilities. The perfect classifier would have only 0 and 1 probabilities, but these would not be able to match historical default rates. They simply represent the probability of belonging to a class with the switch threshold at 0.5, they are not predicted default rates.  Fortunately, it is possible to adjust the probability distribution in order to better match the actual distribution observed in the data, without losing predictive power. This adjustment is referred to as \emph{calibration}. In particular, calibrating a classifier consists in fitting a regressor (known as \emph{calibrator}) which maps the output of the classifier $f_i$ to a calibrated probability $p(y_i=1|f_i)$ in $\left[0, 1\right]$. \\Taking inspiration from the hybrid model structure proposed by Xinran He et all \cite{he2014practical}, based on the concatenation of boosted decision trees and of a probabilistic sparse classifier, we calibrated the output probabilities of the LGBM classifier by fitting a L$2$-Regularized Logistic Regression (LR) (where the inverse of regularization strength, i.e. the $c$ parameter, has been optimized on the out-of-time sample of the training-set) on its one-hot encoded leaf assignments. \\We first fitted the LGBM on the stratified test-set we left aside for the classification task, as the sample used to train the calibrator should not be used to train the target classifier. We treated the output of each individual tree of the LGBM classifier as a categorical feature that takes as value the index of the leaf an instance ends up falling in. We applied one-hot encoding to have dummies indicating leaf assignments on which fitting the LR model. Finally we tested the calibrator on the train-set previously used for LGBM  training phase. This methodology of using an intermediate result and change the output from classification to regression is analogous to what is currently known as Transfer Learning in the Deep Neural Network world, where the final neural net layer is removed and substituted with a novel output. The main advantage of this method is in preserving all the inner complex feature engineering that the system learned in the original training task and transferring it to a different problem, in our specific case the prediction of actual default rates.

\subsection{Rating attribution via genetic algorithm}

A robust default classification system, able to meet both supervisory requirements and internal banking usage, provides a way to represent the internally calibrated probability of default to a rating system, in which each PD bucket is matched to a rating grade. In order to calibrate our own rating system, the refitted default probability has been split into $9$ groups (corresponding to $9$ different rating classes) by means of a genetic algorithm known \emph{Differential Evolution}\cite{differencial_evolution}.\\The algorithmic task of calibrating a rating system can be stated as an optimization problem, as it tries to minimize the Brier Score, maximise the similarities among elements of the same groups (the so called \emph{cohesion} i.e. the items in a cluster should be as similar as possible), minimise the dissimilarities between different groups (the so called \emph{separation} i.e. any two clusters should be as distinct as possible in terms of similarity of items), ensure both PD monotonicity (i.e. lower default rates have to correspond to low rating grades and vice versa) and have an acceptable cluster size  (i.e. each cluster has to include a fraction of the total population that has to be roughly homogeneous among them). Among the partitioning clustering algorithms, Genetic Algorithms (GA) are stochastic search heuristic inspired by the concepts of Darwinian evolution and genetics. They are based on the idea of creating a population of candidate solutions to an optimization problem, which is iteratively refined by alteration (mutation) and selection of good solutions for the next iteration. Candidate solutions are selected according to a so-called fitness function, which evaluates their quality in respect to the optimization problem. In the case of Differential Evolution (DE) algorithms the candidate solutions are linear combinations of existing solutions. In the end, the best individual of the population is returned. This individual represents the best solution discovered by the algorithm.

\section{Results}
\label{section:results}

The metric used to evaluate the model performance is the AUROC or AUC-ROC score (\emph{Area Under the Receiver Operating Characteristics}). In particular, ROC is a probability curve and AUC represents the degree of separability. This measure tells how much a model is capable of distinguishing between classes: for an excellent model it is near to $1$, for a poor model it near to $0$. The ROC curve is constructed by evaluating the fraction of "true positives"(\emph{tpr} or \emph{True Positive Rate}) and "false positives" (\emph{fpr} or \emph{False Positive Rate}) for different threshold values. In details, \emph{tpr}, also known as \emph{Recall} or \emph{Sensitivity}, is defined in \Cref{eqn:tpr} as the number of items correctly identified as positive out of total true positives:

\begin{equation}\label{eqn:tpr}
tpr = \frac{\text{TP}}{\text{TP}+\text{FN}}
\end{equation}

where TP is the number of true positives and FN is the number of false negatives. The \emph{fpr}, also known as \emph{Type I Error}, is defined in \Cref{eqn:fpr} as the number of items wrongly identified as positive out of total true negatives:

\begin{equation}\label{eqn:fpr}
fpr = \frac{\text{FP}}{\text{FP}+\text{TN}}
\end{equation}

where FP is the number of false positive and TN is the number of true negatives.

Prediction results are then summarized into a \emph{confusion matrix} which counts the number of correct and incorrect predictions made by the classifier. A threshold is applied to the cut-off point in probability between the positive and negative classes, which for the default classifier has been set at $0.5$. However, a trade-off exists between \emph{tpr} and \emph{fpr}, such that changing the threshold of classification will change the balance of predictions towards improving the \emph{True Positive Rate} at the expense of \emph{False Positive Rate}, or vice versa.

The metric used to evaluate the performance of internally calibrated PD prediction is the \emph{Brier score} (\text{BS}), i.e. a way to verify the accuracy of a probability forecast in term of “distance” between the actual results. The most common formulation of the \emph{Brier score} is \emph{mean squared error}:
\begin{equation}\label{eqn:brier}
\text{BS} = \frac{1}{\text{N}}\sum_{t=1}^N (f_t - o_t)^2
\end{equation}

in which $f_{t}$ is the forecast probability, $o_{t}$ the actual outcome of the event at instance $t$ and N is the number of forecasting instances. The best possible Brier score is $0$, for total accuracy, the lowest possible score is $1$, which mean the forecast was wholly inaccurate.

Note that, all the metrics described so far have been calculated on the \emph{Test-set} obtained by splitting the dataset along the financial statement year\footnote{Train-set spans from $2011$ to $2016$, Test-set covers $2017$}.

\subsection{Default Classification}

\begin{figure}[h!]
     \centering
     %\vspace{10pt}
	\includegraphics[scale=0.7]{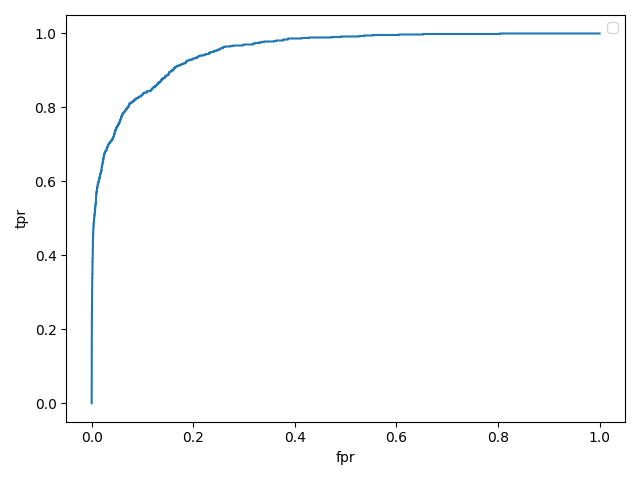}
     \caption{ROC curve for Light-GBM classifier}
     \label{fig:class_ROC}
\end{figure}

\begin{figure}[h!]
     \centering
     %\vspace{10pt}
	\includegraphics[scale=0.7]{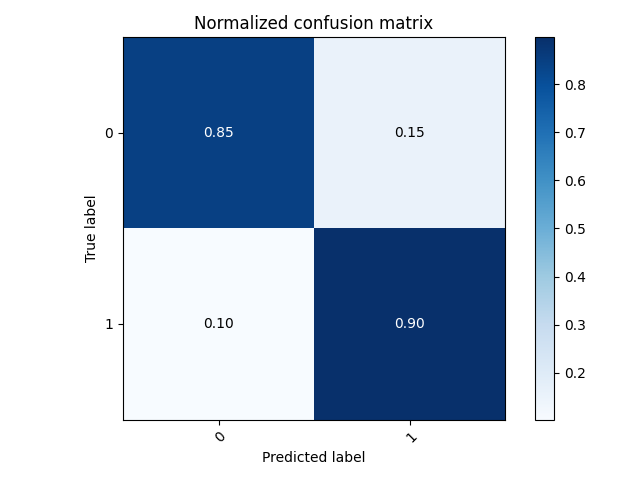}
     \caption{Normalized confusion matrix with $50\%$ threshold for Light-GBM classifier}
     \label{fig:class_confusion_matrix}
\end{figure}

We obtained an high performance corresponding to an AUROC $= 95.0\%$ (see \autoref{fig:class_ROC}) and summarized in the normalized confusion matrix of \autoref{fig:class_confusion_matrix}. In highly unbalanced datasets, usually the confusion matrix is skewed towards predicting well only the majority class, producing unsatisfactory performances in the minority class, even if the event of misclassification for the minority class is the event business usually try to minimize the most. The distorsion in the default probability space and an accurate choice of feature selection and hyperparameters created a system able to effectively discriminate events in the minority class, reducing the occurrence of false negatives.

%\vspace{10pt}
\FloatBarrier

%\vspace{10pt}
\FloatBarrier

\subsection{PD refitting}

\begin{figure}[h!]
 \centering
 \subfloat[]{\includegraphics[scale=0.4]{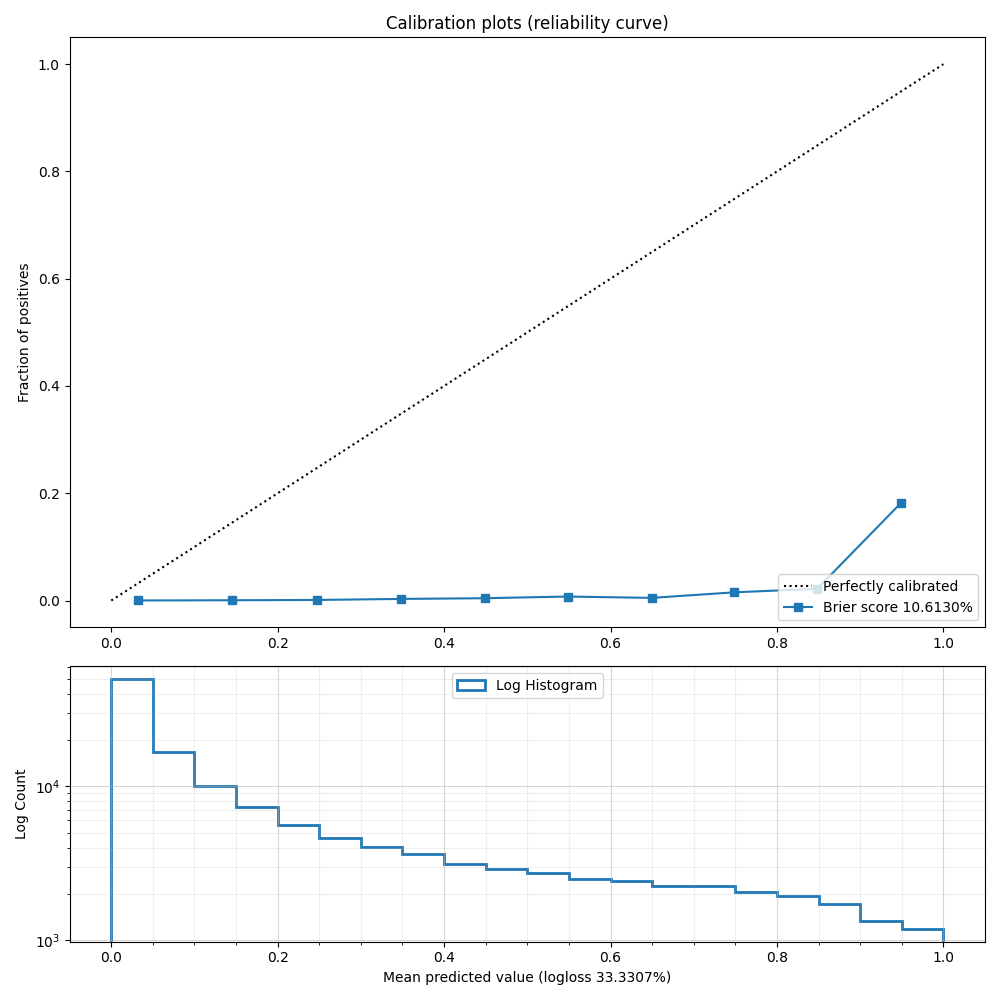}\label{fig:a}}\\
 \subfloat[]{\includegraphics[scale=0.4]{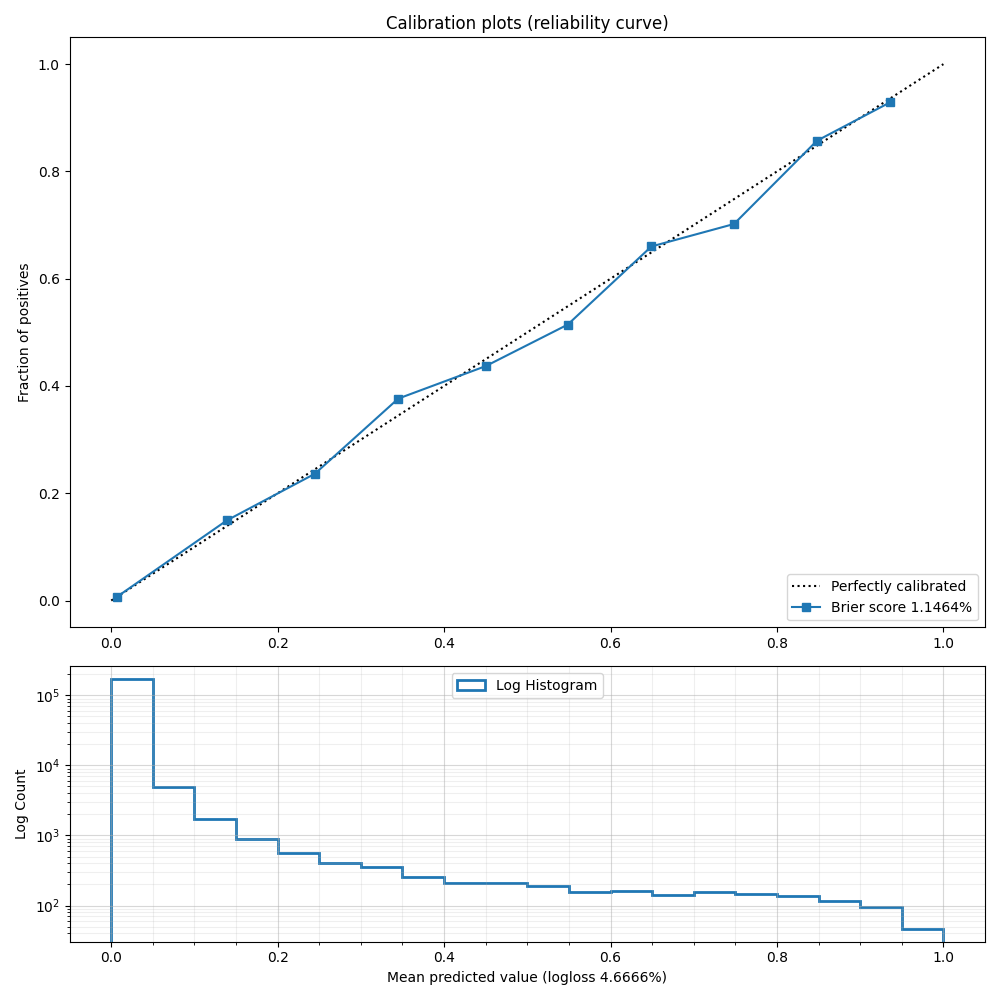}\label{fig:b}}
 \caption{Calibration plots and log-scaled histograms of forecast probability before (\Cref{fig:a}) and after (\Cref{fig:b}) refitting. Accuracy of predicted probabilities is expressed in term of \emph{log loss} measure}
 \label{fig:calibration_plot}
\end{figure}

Default probabilities forecast before and after refitting procedure are summarized in the \emph{calibration plots} (also called \emph{reliability curves}) of \autoref{fig:calibration_plot}, which allow checking if the predicted probabilities produced by the model are well calibrated. Specifically, a calibration plot consists of a line plot of the relative observed frequency (y-axis) versus the predicted probabilities (x-axis)\footnote{In details, the predicted probabilities are divided up into a fixed number of buckets along the x-axis. The number of target events (i.e. the occurrence of 1-year default) are then counted for each bin (i.e. the relative observed frequency). Finally, the counts are normalized and the results are plotted as a line plot}. A perfect classifier would produce only 0 and 1 predictions but would not be able to forecast actual default rates. A perfect actual default rate model would produce reliability diagrams as close as possible to the main diagonal from the bottom left to the top right of the plot. The refitting procedure maps the perfect classifier to a reliable default rate predictor.

%\FloatBarrier

\begin{figure}[h!]
     \centering
     %\vspace{10pt}
	\includegraphics[scale=0.7]{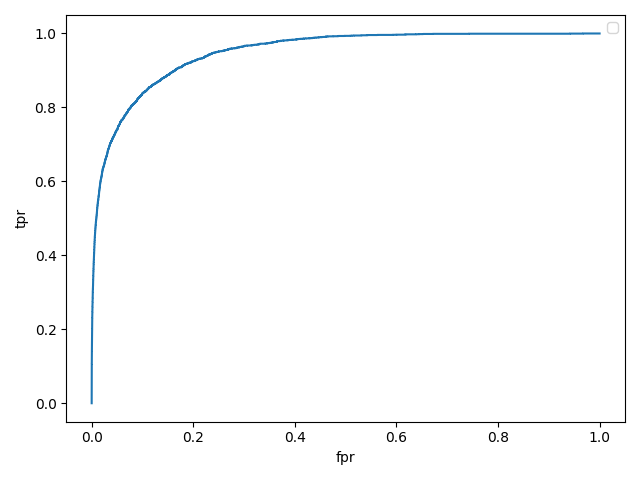}
     \caption{ROC curve for the calibrated classifier}
     \label{fig:pd_ROC}
\end{figure}

\begin{figure}[h!]
      \centering
      %\vspace{10pt}
 	\includegraphics[scale=0.7]{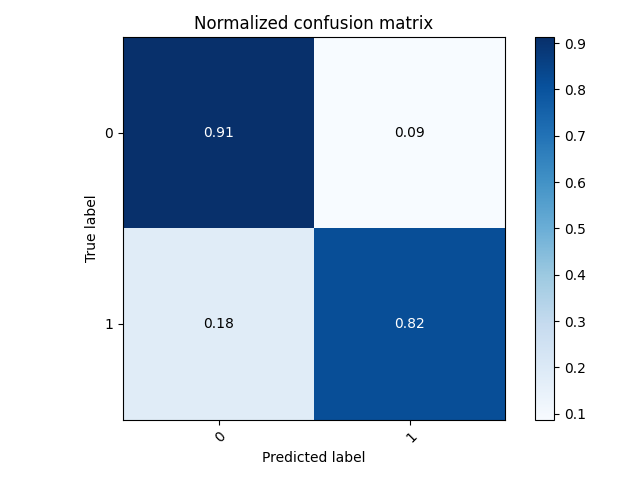}
      \caption{Normalized confusion matrix with optimized threshold for the calibrated classifier}
      \label{fig:pd_confusion_matrix}
 \end{figure}

The refitting procedure left AUROC score of the model unchanged (AUROC $= 95.0\%$); the calibration performance is evaluated with the Brier score ($BS = 1.2\%$\footnote{The closer the Brier score is to zero the better is the forecast of default probabilities}); classification results are summarized in the normalized confusion matrix reported in \autoref{fig:pd_confusion_matrix}, where the threshold to the cut-off point between the positive and negative classes has been optimized on the \autoref{fig:pd_ROC}.

%\vspace{10pt}
%\FloatBarrier

 %\vspace{10pt}
\FloatBarrier

\subsection{PD clustering}

Among the several common statistical tests that can be performed to validate the assignment of a probability of default to a certain rating grade, two approaches have been used: the \emph{one-sided Binomial test} and the \emph{Extended Traffic-Light Approach}.

The \emph{Binomial Test} is one of the most popular single-grade single-period\footnote{usually one year} test performed for rating system validation. For a certain rating grade $k \in \left\{1, \dots, K\right\}$, being $K$ the number of rating classes\footnote{9 in our case}, we made the assumptions that default events are independent within the grade $k$ and could be modelled as a binomially distributed random variable $X$ with size parameter $N_{k}$ and "\emph{success}" probability $PD_{k}$. Thus, we can assess the correctness of the PD forecast by testing the null hypothesis H$_{0}$, where:
\begin{itemize}
\item H$_{0}$: the actual default rate is less than or equal to the forecast default rate given by the PD;
\end{itemize}

The null hypothesis H$_{0}$ is rejected at a confidence level $\alpha$ in case the number of observed defaults $d$ per rating grade is greater than or equal to the critical value, as reported in \autoref{eqn:binomial}:

\begin{equation}\label{eqn:binomial}
d_{\alpha} = \min{\left\{d: \sum_{j=d}^{N_{k}} \binom{N_{k}}{j}PD^{j}_{k}(1-PD^{j}_{k})^{N_{k}-j} \leq 1-\alpha\right\}}
\end{equation}

The \emph{Extended Traffic Light Approach} is a novel technique for default probability validation, first adopted by Tasche (2003)\cite{tasche2003traffic}. The implementation used in this section refers to a heuristic approach proposed by Blochwitz et al. (2005) \cite{blochwitz2005reconsidering} which is based on the estimation of a relative distance between observed default rates and forecast probabilities of default, under the key assumption of binomially distributed default events. Four coloured zones, \emph{Green}, \emph{Yellow}, \emph{Orange}, \emph{Red}, are established to analyse the deviation of forecasts and actual realizations. In details: if the result of the validation assessment lies in the \emph{Green} zone there is no obvious contradiction between forecast and realized default rate; the \emph{Yellow} and \emph{Orange} lights indicate that the realized default rate is not compatible with the PD forecast, however, the difference of realized rate and forecast is still in the range of usual statistical fluctuations; and last \emph{red} traffic light indicates a wrong forecast of default probability. The boundaries between the afore mentioned light-zones are summarized in \autoref{eq:traffic_light}:

\begin{equation}\label{eq:traffic_light}
\begin{cases}
\mbox{Green} & p_{k} < PD_{k}\\
\mbox{Yellow} & PD_{k} \leqslant p_{k} < PD_{k}+ K^{y}\sigma(PD_{k}, N_{k})\\
\mbox{Orange} & PD_{k}+ K^{y}\sigma(PD_{k}, N_{k}) \leqslant p_{k} < PD_{k}+ K^{0}\sigma(PD_{k}, N_{k})\\
\mbox{Red} & PD_{k}+ K^{0}\sigma(PD_{k}, N_{k}) \leqslant p_{k}
\end{cases}
\end{equation}

where $\sigma(PD_{k}, N_{k}) = \sqrt{PD_{k}(1-PD_{k})/N_{k}}$. The parameters $K^{y}$ and $K^{0}$ play a major role in the validation assessment, so have to be tuned carefully. A proper choice based on practical considerations is setting $K^{y}= 0.84$ and $K^{0}= 1.44$, which corresponds to a probability of observing green of $0.5$, observing yellow with $0.3$, orange with $0.15$ and red with $0.05$.

The results of the application of the above mentioned statistical test are summarized in \autoref{tab:rating}.

\begin{table}[h!]
\begin{tabular}{c|c|c|c|c|c}
\hline
\textbf{\begin{tabular}[c]{@{}c@{}}Rating\\class\end{tabular}} &
  \textbf{\begin{tabular}[c]{@{}c@{}}PD Bins\\(\%)\end{tabular}} &
  \textbf{\begin{tabular}[c]{@{}c@{}}Rating Class\\PD (\%)\end{tabular}} &
  \textbf{\begin{tabular}[c]{@{}c@{}}Out-of-sample\\Default Rate\\(\%)\end{tabular}} &
  \textbf{\begin{tabular}[c]{@{}c@{}}One-sided\\Binomial Test\end{tabular}} &
  \textbf{\begin{tabular}[c]{@{}c@{}}Extended\\Traffic Light\\Approach\end{tabular}} \\ \hline
AAA & {[}0.00, 0.005) & 0.03  & 0.00  & Passed & Green \\
AA  & {[}0.05, 0.42)  & 0.24  & 0.03  & Passed & Green \\
A   & {[}0.42, 0.55)  & 0.48  & 0.08  & Passed & Green \\
BBB & {[}0.55, 0.74)  & 0.64  & 0.21  & Passed & Green \\
BB  & {[}0.74, 1.00)  & 0.87  & 0.40  & Passed & Green \\
B   & {[}1.00, 1.42)  & 1.21  & 0.83  & Passed & Green \\
CCC & {[}1.42, 2.12)  & 1.77  & 1.29  & Passed & Green \\
CC  & {[}2.12, 9.03)  & 5.57  & 5.06  & Passed & Green \\
C   & {[}9.03,100)    & 54.52 & 33.77 & Passed & Green \\ \hline
\end{tabular}
\caption{Internally calibrated PD clustering into 9 rating classes. Despite being borrowed from S\&P rating scales, the labels are assigned to a PD calibrated on an internal dataset (the one used during the training-phase) and does not correspond to any rating agencies PD}
\label{tab:rating}
\end{table}
\FloatBarrier

\section{Model explainability}
\label{section:explain}

Machine learning models which operate in higher dimensions than cannot be directly visualized by human mind are often referred as "\emph{black boxes}", in the sense that high model performance is often achieved on the detriment of output explainability, leading users not to understand the logic behind model predictions. Even greater attention to model interpretability has led to the development of several methods to provide an explanation to machine learning outputs, both in term of global and local interpretability. In the first case, the goal is being able to explain and understand model decisions based on conditional interactions between the dependent variable (i.e. target) and the independent features on the entire dataset. In the latter case, the aim is to understand model output for a single prediction by looking at a local subregion in the feature space around that instance. \\Two popular approaches described hereafter in this section are SHAP and LIME, which explore and leverage the property of \emph{local explainability} to build surrogate models which are able to interpret the output of any machine learning models. The technique upon which these algorithms are based is slightly tweaking the input and modelling the changes in prediction by means of surrogate agnostic models. In particular, SHAP measures how much each feature in our model contributes, either positively or negatively, to each prediction, in term of difference between the actual prediction and its expected value. LIME builds sparse linear models around each prediction to explain how the black box model works in that local vicinity.

\subsection{SHAP}

SHAP, which stands for (\textsl{SH}apley \textsl{A}dditive ex\textsl{P}lanation) \cite{lundberg2017unified}, is a novel approach for model explainability which exploits the idea of \emph{Shapley regression value}\footnote{The technical definition of Shapley value is the average marginal contribution of a feature value over all possible coalitions} to model feature influence scoring. SHAP values quantify the magnitude and direction (positive or negative) of a feature's effect on a prediction via an additive feature attribution method. In simple words, SHAP builds model explanations by asking, for each prediction $i$ and feature $j$, how $i$ changes when $j$ is removed from the model. Since SHAP considers all possible predictions for an instance using all possible combinations of feature inputs, it can guarantee both consistency and local accuracy. More in details, SHAP method computes Shapley values from \emph{coalitional game theory}. The feature values of a data instance act as players\footnote{note that a player can be an individual feature value or  a group of feature values} in a coalition: Shapley values suggest how to fairly distribute the “\emph{payout}” (i.e. the prediction) among the features.

\paragraph{SHAP summary plot}

As reported in \Cref{fig: summary_plot}, it combines feature importance with feature effects to measure the global impact of features on the model. For each feature shown in the y-axis, and ordered according to their importance, each point on the plot represents the Shapley value (reported along the x-axis) for a given prediction. The color of each point represents the impact of the feature on model output from low (i.e. blue) to high (i.e. red). Overlapping points are littered in y-axis direction, so we get a sense of the distribution of the Shapley values per feature.

%\newpage
\begin{figure}[h!]
     \centering
	\includegraphics[scale=0.6]{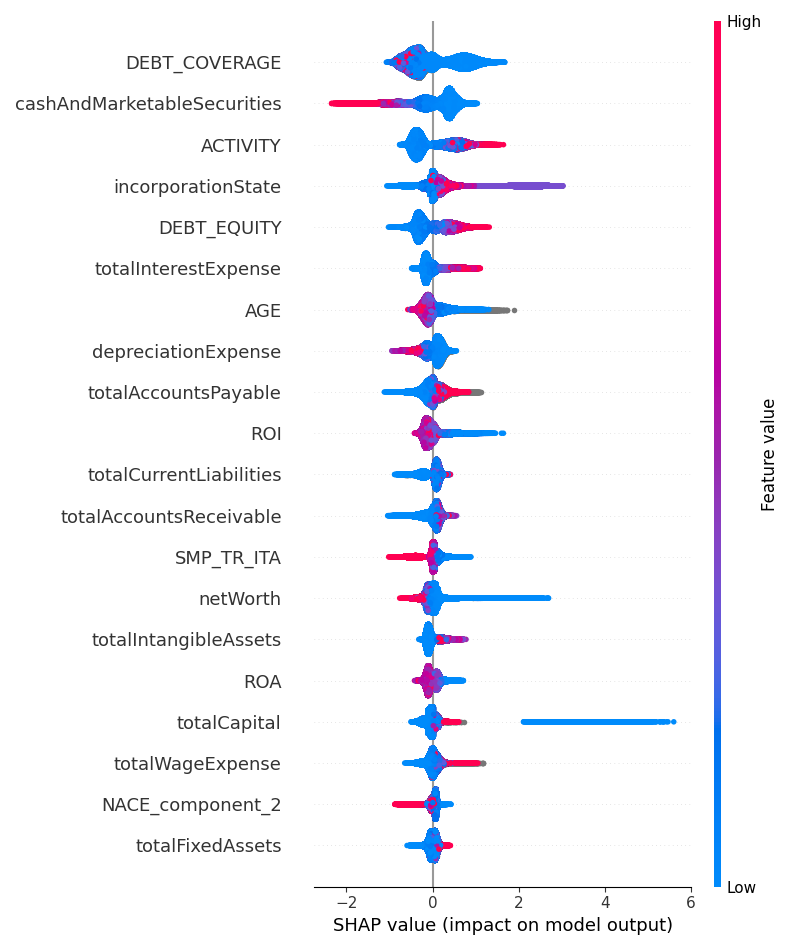}
     \caption{SHAP summary plot for Light-GBM classifier. The details of model's feature description are reported in Appendix \ref{appendix}}
     \label{fig: summary_plot}
\end{figure}
\FloatBarrier

%%Commento al grafico
%It appears that the most important financial ratio predictor for the default probability of a company is .... followed by the availability of working capital and Interest Expense Coverage. In essence ... may assure the viability of a business. In addition the economic climate, seem to play an additional important role in business viability since the Economic sentiment indicator and the Consumer confidence indicator are rendered important in the model whereas other widely employed factors such as ... seem not to be predominant.

\paragraph{SHAP dependence plot}

It is a scatter plot that shows the effect a single feature has on the model predictions. In particular, each dot represents a single prediction where the feature value is on the x-axis and its SHAP value, representing how much knowing that feature's value changes the output of the model for that sample's prediction, on the y-axis. The color corresponds to a second feature that may have an interaction effect with the plotted feature. If an interaction effect is present it will show up as a distinct vertical pattern of colouring.

%\newpage
\begin{figure}[h!]
 \centering
 \subfloat[]{
 \includegraphics[scale=0.35]{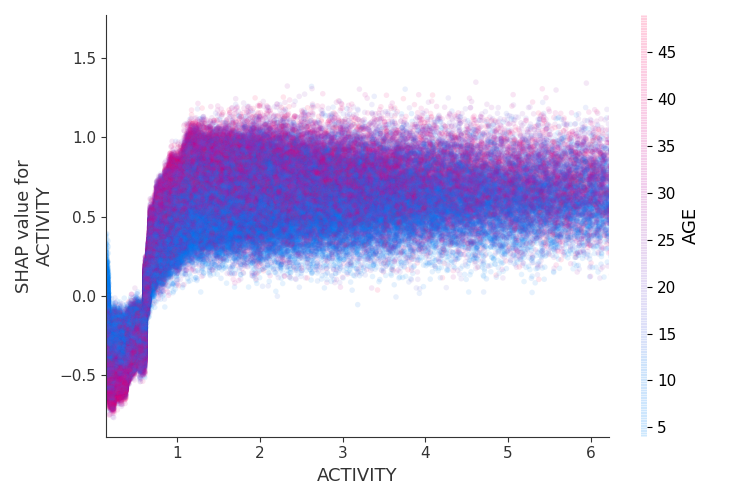}\label{fig:a}}
 \subfloat[]{\includegraphics[scale=0.35]{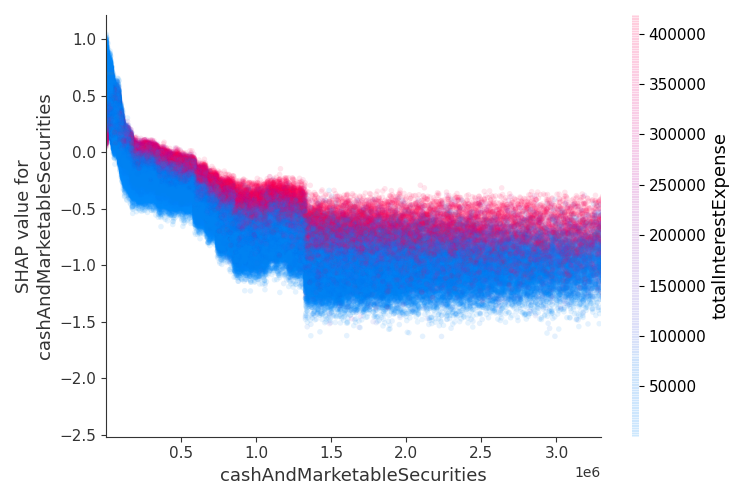}\label{fig:b}}\\
 \subfloat[]{\includegraphics[scale=0.35]{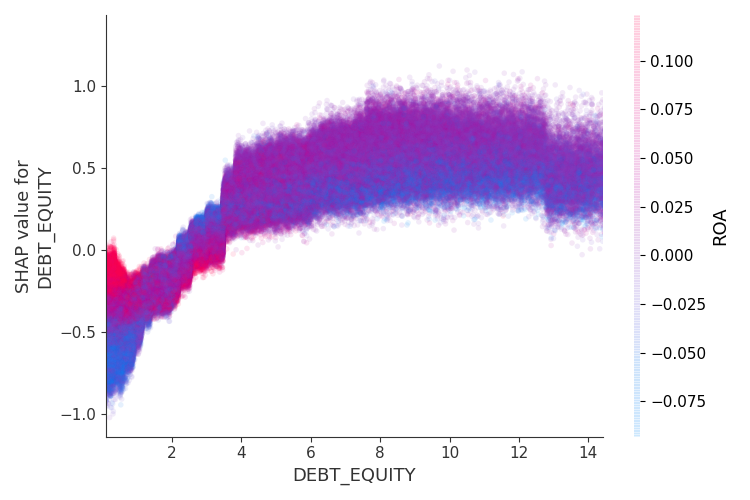}\label{fig:c}}
 \subfloat[]{\includegraphics[scale=0.35]{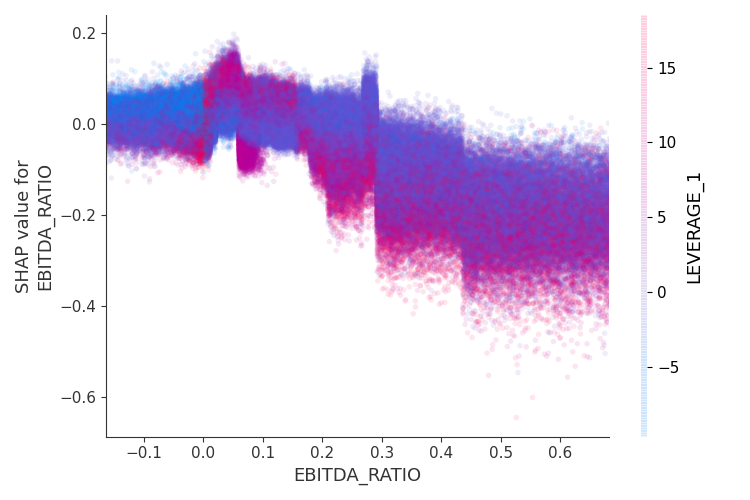}\label{fig:d}}\\
 \subfloat[]{\includegraphics[scale=0.35]{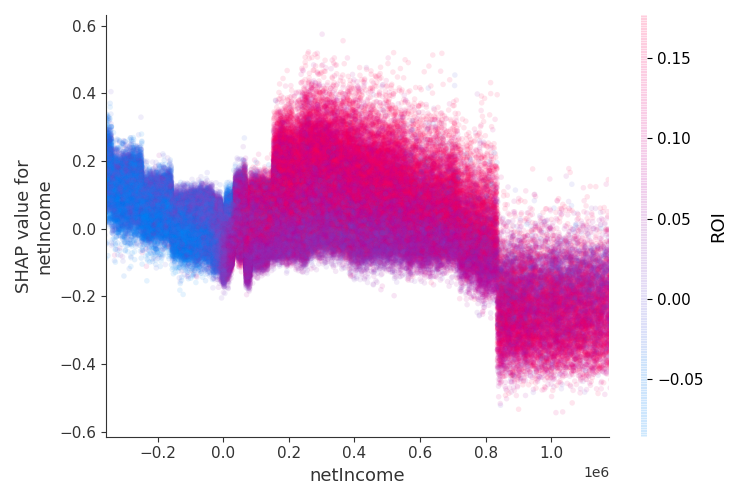}\label{fig:e}}
 \subfloat[]{\includegraphics[scale=0.35]{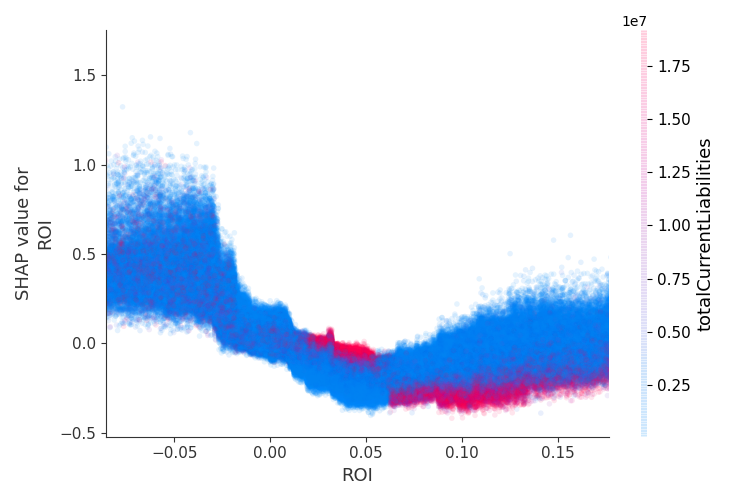}\label{fig:f}}\\
 \subfloat[]{\includegraphics[scale=0.35]{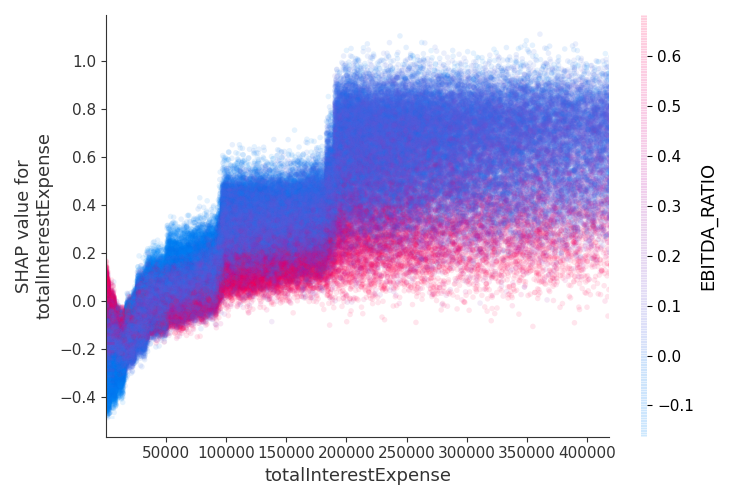}\label{fig:g}}
 \caption{SHAP dependency plots for ACTIVITY in~\Cref{fig:a}, cashAndMarketableSecurities in~\Cref{fig:b}, DEBT$\_$EQUITY in~\Cref{fig:c}, EBITDA$\_$RATIO in~\Cref{fig:d}, netIncome in~\Cref{fig:e}, ROI in~\Cref{fig:f}, totalInterestExpense in~\Cref{fig:g}. The details of model's feature description are reported in Appendix \ref{appendix}}
 \label{some-label}
\end{figure}
%\FloatBarrier

\paragraph{SHAP waterfall plot}

\begin{figure}[h!]
     \centering
	\includegraphics[scale=0.6]{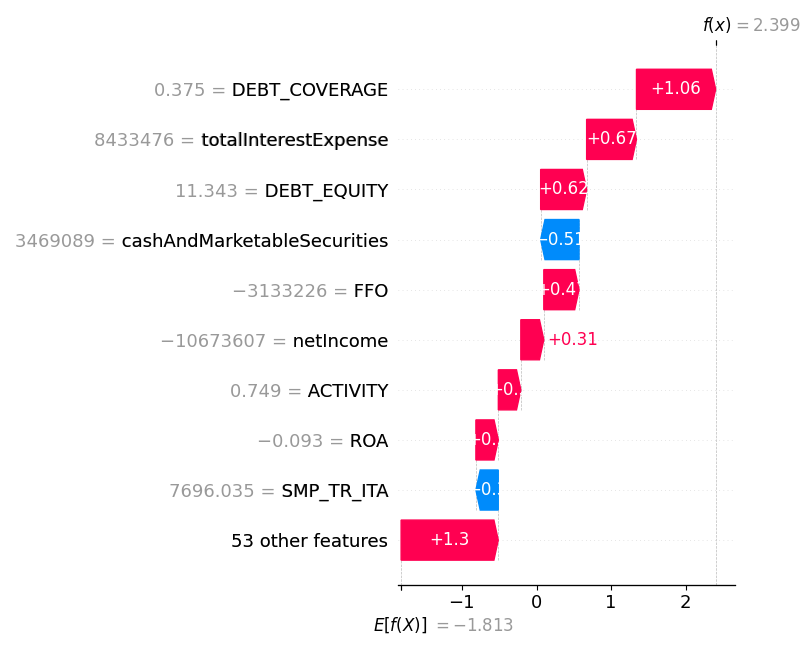}
     \caption{SHAP waterfall plot %for Light-GBM classifier.
     The details of model's feature description are reported in Appendix \ref{appendix}}
     \label{fig: waterfall_plot}
\end{figure}

The waterfall plot reported in \autoref{fig: waterfall_plot} is designed to display how the SHAP values of each feature move the model output from our prior expectation under the background data distribution ($E[(f(X)]$), to the final model prediction ($f(X)$) given the evidence of all the features. Features are sorted by the magnitude of their SHAP values with the smallest magnitude features grouped together at the bottom of the plot. The color of each row represents the impact of the feature on model output from low (i.e. blue) to high (i.e. red).

%\newpage
\FloatBarrier

\subsection{Lime}

LIME, \textsl{L}ocal \textsl{I}nterpretable \textsl{M}odel-agnostic \textsl{E}xplanations, is a novel technique that explains the predictions of any classifier in an interpretable and faithful manner, by learning an interpretable model locally around the prediction\cite{ribeiro2016should}. Behind the workings of LIME lies the assumption that every complex model is linear on a local scale, so it is possible to fit a simple model around a single observation that will mimic how the global model behaves at that locality. The output of LIME is a list of explanations, reflecting the contribution of each feature to the prediction of a data sample, allowing to determine which feature changes will have most impact on the prediction. Note that, LIME has the desirable property of additivity, i.e. the sum of the individual impact is equal to the total impact. Results for a prediction are summarized in \autoref{fig: summary_plot}.

%\newpage
\begin{figure}[h!]
     \centering
	\includegraphics[scale=0.9]{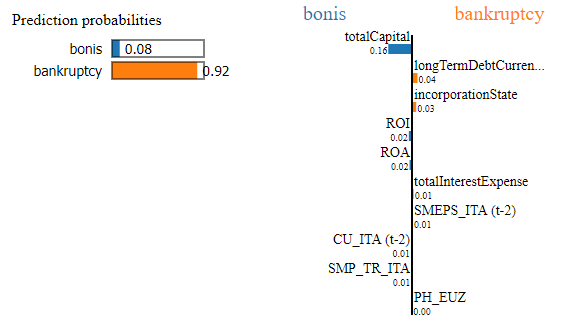}
     \caption{Lime local explanation for a prediction from Light-GBM classifier. The details of model's feature description are reported in Appendix \ref{appendix}}
     \label{fig: summary_plot}
\end{figure}
\FloatBarrier

%%%%%%%%%%%%%%%%%%%%%%%%%%%%%%%%%%%%%%%%%%%%%%%%%%%%%%%%%%%%%%%%%%%%%%%%%%%%%%%%%%%%%%%%%

\section{Conclusions}
\label{section:conclusions}

Starting from Moody's dataset of historical balancesheets, bankruptcy statuses and macroeconomic variables we have built three models: a classifier, a default probability model and a rating system. By leveraging on modern techniques in both data processing and parameter calibration we have reached state of the art results. The three models show stunning out of sample performances allowing for an intensive usage in risk averse businesses where the occurrence of false negatives can dramatically harm the firm itself. The explainability layers via Shap and Lime give a set of extra tools to increase the confidence in the model and help in understanding the main features determining a specific result. These information can be leveraged by the analyst to understand how to reduce the bankruptcy probability of a specific firm or to get insight in which balance sheet fields need to be improved to increase the rating, therefore providing a business instrument to actively manage clients and structured finance deals.

\section*{Acknowledgements}

We are grateful to Corrado Passera for encouraging our research.

%%%%%%%%%%%%%%%%%%%%%%%%%BIBLIOGRAPHY%%%%%%%%%%%%%%%%%%%%%%%%%%%%%%%%%%%%%%%%%%%
%\bibliographystyle{plain}
\bibliographystyle{ieeetr}
\bibliography{mybib}

%%%%%%%%%%%%%%%%%%%%%%%%%APPENDIX%%%%%%%%%%%%%%%%%%%%%%%%%%%%%
\newpage
\newpage

\begin{appendices}
\section{Model's Features descriptions}\label{appendix}

In this section the details of selected features, upon which the model have been trained, are reported.

\subsection{Balancesheet index descriptions}\label{appendix: balance}

\begin{center}
\begin{longtable}{| L{4cm} | L{11cm} |}

    \hline
    \hline
    \textbf{Code} & \textbf{Definition} \\\hline \endhead
    \hline
    cashAndMarketableSecurities  & Cash and marketable securities. \\
    \hline
    depreciationExpense  & The depreciation expense for current period.\\
    \hline
    ebitda  & Earnings before interest, taxes, depreciation and amortization before extraordinary items. \\
    \hline
    entityConsolidationType  & For companies with subsidiaries, i\\
    \hline
    incorporationRegion  & The entity's incorporation region.\\
    \hline
    incorporationState  & The entity's incorporation province or administrative division where the  entity has a legal representation. \\
     \hline
    longTermDebtCurrentMaturities  & The current maturities of long-term debt, principal payments due within 12 months. \\
    \hline
    netIncome  & Net income is the total period-end earnings. \\
    \hline
    netWorth  & Net worth is the sum of all equity items, including retained earnings and other equity.\\
    \hline
    payableToTrade  & Accounts payable to regular trade accounts.\\
    \hline
    receivableFromTrade  & Accounts receivable from trade.\\
    \hline
    retainedEarnings  & Retained earnings.\\
    \hline
    tangibleNetWorth  & Defined as the difference between netWorth and totalIntangibleAssets.\\
    \hline
    totalAccountsPayable  & The sum of accounts payable.\\
    \hline
    totalAccountsReceivable  & The total accounts receivable, net of any provision from loss.\\
    \hline
    totalAmortizationAndDepreciaton  & The sum of amortization and depreciation expense for current period.\\
    \hline
    totalAssets  & The total assets of the borrower which is the sum of the current assets and  non-current assets.\\
    \hline
    totalCapital  & Total subscribed and share capital.\\
    \hline
    totalCapital\_and\_totalLiabilities  & Defined as the sum of totalLiabilities and totalCapital\\
    \hline
    totalCurrentAssets  & The sum of all current assets. \\
    \hline
    totalCurrentLiabilities  & The sum of all current liabilities.\\
    \hline
    totalFixedAssets  & Total fixed assets are the Gross Fixed Assets less Accumulated Depreciation.\\
    \hline
    totalIntangibleAssets  & Total intangible assets.\\
    \hline
    totalInterestExpense  & The total interest expense is any gross interest expense generated from short-term, long-term, subordinated or related debt. \\
    \hline
    totalInventory  & The sum of all the inventories.\\
    \hline
    totalLiabilities  & The sum of Total Current Liabilities and  Total non-current liabilities.\\
    \hline
    totalLongTermDebt  & The amount due to financial and other institutions after 12 months.\\
    \hline
    totalOperatingExpense  & The sum of all operating expenses.\\
    \hline
    totalOperatingProfit  & The Gross Profit less Total Operating Expense\\
    \hline
    totalProvisions  & Total provisions for pensions, taxes, etc.\\
    \hline
    totalSales  & Total sales.\\
    \hline
    totalWageExpense  & The total wage expense.\\
    \hline
    workingCapital  & Defined as the sum of receivableFromTrade, totalAccountsReceivable and totalInventory minus payableToTrade and totalAccountsPayable.\\

\hline
\hline
\end{longtable}
\end{center}

\subsection{KPIs descriptions}\label{appendix: kpis}

\begin{equation}
\text{ACID} = \frac{\text{cashAndMarketableSecurities} + \text{totalAccountsReceivable}}{\text{totalCurrentLiabilities}}
\end{equation}

\begin{equation}
\text{ACTIVITY} = \frac{\text{totalCurrentLiabilities}}{\text{totalSales}}
\end{equation}

\begin{equation}
\text{AGE} = \frac{\text{financialStatementDate} -  \text{incorporationDate}}{365.0}
\end{equation}

being "financialStatementDate" the date of the financial statement and "incorporationDate" the date in which the entity was incorporated

\begin{equation}
\text{ASSET\_TURNOVER} = \frac{\text{totalSales}}{\text{totalAssets}}
\end{equation}

\begin{equation}
\text{CURRENT\_RATIO} = \frac{\text{totalCurrentAssets}}{\text{totalCurrentLiabilities}}
\end{equation}

\begin{equation}
\text{DEBT\_COVERAGE} = \frac{\text{ebitda}}{\text{totalInterestExpense}}
\end{equation}

\begin{equation}
\text{DEBT\_EQUITY} = \frac{\text{totalLiabilities}}{\text{netWorth}}
\end{equation}

\begin{equation}
\text{EBITDA\_RATIO} = \frac{\text{ebitda}}{\text{totalSales}}
\end{equation}

\begin{equation}
\text{FFO} = \text{netIncome} + \text{totalAmortizationAndDepreciaton} + \text{depreciationExpense}
\end{equation}

\begin{equation}
\text{IND\_ROTA} = \frac{\text{workingCapital}}{\text{totalSales}}
\end{equation}

\begin{equation}
\text{IND\_STRUTT} = \frac{\text{PFN}}{\text{netWorth}}
\end{equation}

\begin{equation}
\text{INVENTORY\_TURNOVER} =  \frac{\text{totalInventory}}{\text{totalSales}}
\end{equation}

\begin{equation}
\text{LEVERAGE\_1} = \frac{\text{PFN}}{\text{ebitda}}
\end{equation}

\begin{equation}
\text{LEVERAGE\_2} = \frac{\text{FFO}}{\text{PFN}}
\end{equation}

\begin{equation}
\text{LONG-TERM-DEBT\_EQUITY} = \frac{\text{totalLongTermDebt}}{\text{netWorth}}
\end{equation}

\begin{equation}
\text{NETINCOME\_RATIO} = \frac{\text{netIncome}}{\text{totalSales}}
\end{equation}

\begin{equation}
\text{PFN} = \text{totalLongTermDebt} + \text{longTermDebtCurrentMaturities}
        - \text{cashAndMarketableSecurities}
\end{equation}

\begin{equation}
\text{ROA} =  \frac{\text{netIncome}}{\text{totalAssets}}
\end{equation}

\begin{equation}
\text{ROE} = \frac{\text{netIncome}}{\text{netWorth}}
\end{equation}

\begin{equation}
\text{ROI} = \frac{\text{totalOperatingProfit}}{\text{totalAssets}}
\end{equation}

\begin{equation}
\text{SHORT-TERM-DEBT\_EQUITY} = \frac{\text{totalCurrentLiabilities}}{\text{netWorth}}
\end{equation}

\subsection{Macro-economic factors descriptions}\label{appendix: macro}

\begin{center}
\begin{longtable}{| L{4cm} | L{4cm} | L{7cm} |}
    \multicolumn{3}{c}{\textbf{Country specific indicators}} \\
    \hline
    \hline
    \textbf{Code} & \textbf{Name} & \textbf{Definition} \\\hline \endhead
    \hline
    C & Consumption, private, real & The volume of goods and services consumed by households and non-profit institutions serving households. \\
    \hline
CD & Durable goods & The volume of real personal consumption expenditures. \\
\hline
CREDR & Credit rating, average & The sovereign risk rating, based on the average of the sovereign ratings provided by Moody’s, S\&P and Fitch. \\
\hline
CU & Capacity utilisation & A measure of the extent to which the productive capacity of a business is being used. \\
\hline
DOMD & Domestic demand, real & The volume of consumption, investment, stockbuilding and government consumption expressed in local currency and at prices of the country's base year. \\
\hline
EE & Employees in employment & Employees in employment. \\
\hline
ET & Employment, total & Employment, total. \\
\hline
GC & Consumption, government, real & The volume of government spending on goods and services. \\
\hline
GDP & GDP, real & The volume of all final goods and services produced within a country in a given period of time. \\
\hline
GDPHEAD & GDP per capita, real, US\$, constant prices & GDP per capita, real, US\$, constant prices. \\
\hline
IF & Investment, total fixed investment, real & Investment, total fixed investment, real. \\
\hline
IP & Industrial production index & The volume of investment in tangible and intangible capital goods, including machinery and equipment, software, and construction. \\
\hline
IPNR & Investment, private sector business, real & The volume of investment in private sector business. \\
\hline
IPRD & Investment, private dwellings, real & The volume of investment in private dwellings. \\
\hline
IS & Stockbuilding, real & The volume of stocks of outputs that are still held by the units that produced them and stocks of products acquired from other units that are intended to be used for intermediate consumption or for resale. \\
\hline
M & Imports, goods \& services, real & The volume of goods and services imports. \\
\hline
MG & Imports, goods, real & The volume of goods imports. \\
\hline
MS & Imports, services, real & The volume of services imports. \\
\hline
PEWFP & GDP, compensation of employees, total, nominal & The values of wages and salaries of employees as a component of GDP. \\
\hline
PH & House price index & Index of house prices. \\
\hline
POIL\$ & Oil price US\$ per toe & Oil price US\$ per toe. \\
\hline
RCB & Interest rate, central bank policy & The rate that is used by central bank to implement or signal its monetary policy stance (expressed as an average). \\
\hline
RCORP$\_$SPREADEOP & Credit spreads, end of period & The difference in yield between two bonds of similar maturity but different credit quality, expressed as end of period value. \\
\hline
RLG & Interest rate, long-term government bond yields & Interest rate, long-term government bond yields. \\
\hline
RS & Retail Sales volume index, excluding automotive & Volume index for retail sales excluding automotive. \\
\hline
RSH & Interest rate, short-term & The 3-month interbank rate. \\
\hline
RSHEOP & Interest rate, short-term, end of period & The 3-month interbank rate for the end of period. \\
\hline
SMEPS & Stockmarket earnings per share & Stockmarket earnings per share calculated as Stockmarket earnings, LCU * 1000 / Stockmarket shares outstanding. \\
\hline
SMP$\_$TR & Share price total return index & Share price total return index. \\
\hline
TFE & Total final expenditure, real & The sum of volumes of consumption, investment, stockbuilding, government consumption and exports. \\
\hline
U & Unemployment & The total number of people without a job, but actively searching for one. \\
\hline
UP & Unemployment rate & The percentage of the labour force that is unemployed at a given date. \\
\hline
X & Exports, goods \& services, real & The volume of goods and services exports expressed in local currency and at the country's base year. \\
\hline
XG & Exports, goods, real & The volume of goods exports expressed in local currency. \\
\hline
XS & Exports, services, real & The volume of services exports expressed in local currency. \\
\hline
\hline
\end{longtable}
\end{center}

\begin{center}
\begin{longtable}{| L{4cm} | L{4cm} | L{7cm} |}
    \multicolumn{3}{c}{\textbf{Eurozone indicators}} \\
    \hline
    \hline
    \textbf{Code} & \textbf{Name} & \textbf{Definition} \\\hline \endhead
	\hline
    PH & House price index & Index of house prices. \\
    \hline
    RCB & Interest rate, central bank policy & The rate that is used by central bank to implement or signal its monetary policy stance (expressed as an average). \\
    \hline
    RCBEOP & Interest rate, central bank policy, end of period & The rate that is used by central bank to implement or signal its monetary policy stance (expressed as an end of period value).\\
    \hline
    REONIA & Interest rate, EONIA & 1-day interbank interest rate for the Eurozone. \\
    \hline
    RLG & Interest rate, long-term government bond yields & Interest rate, long-term government bond yields. \\
    \hline
    RSH & Interest rate, short-term & The 3-month interbank rate.  \\
    \hline
    RSH6M & Interest rate, 6-month & The 6-month interbank rate. \\
    \hline
    RSHEOP & Interest rate, short-term, end of period & The 3-month interbank rate for the end of period. \\
    \hline
    RSWAP2YR & Interest Rate Swap, 2-year & Swap par-rate for 2 years tenor. \\
    \hline
    RSWAP5YR & Interest Rate Swap, 5-year & Swap par-rate for 5 years tenor. \\
    \hline
    RSWAP10YR & Interest Rate Swap, 10-year & Swap par-rate for 10 years tenor. \\
    \hline
    RSWAP30YR & Interest Rate Swap, 30-year & Swap par-rate for 30 years tenor. \\

\hline
\hline
\end{longtable}
\end{center}

\end{appendices}
\end{document}